\documentclass[12pt]{article}
\pdfoutput =1
\usepackage{float} 
\usepackage{amsmath}
\usepackage{slashed}
\usepackage{amsfonts}   
\usepackage{amssymb}

\begin{document}
\date{}
\title{{\bf{\Large Carroll membranes}}}
\author{
 {\bf {\normalsize Dibakar Roychowdhury}$
$\thanks{E-mail:  dibakarphys@gmail.com, dibakarfph@iitr.ac.in}}\\
 {\normalsize  Department of Physics, Indian Institute of Technology Roorkee,}\\
  {\normalsize Roorkee 247667, Uttarakhand, India}
\\[0.3cm]
}

\maketitle
\begin{abstract}
We explore Carroll limit corresponding to M2 as well as M3 branes propagating over 11D supergravity backgrounds in M theory. In the first part of the analysis,
we introduce the membrane Carroll limit associated to M2 branes propagating over M theory supergravity backgrounds.  Considering two specific M2 brane embeddings, we further outline the solutions corresponding to the Hamilton's dynamical equations in the Carroll limit. We further consider the so called \textit{stringy} Carroll limit associated to M2 branes and outline the corresponding solutions to the underlying Hamilton's equations of motion by considering specific M2 brane embeddings over 11D target space geometry. As a further illustration of our analysis, considering the Nambu-Goto action, we show the equivalence between different world-volume descriptions in the Carroll limit of M2 branes.  Finally, considering the \textit{stringy} Carroll limit, we explore the constraint structure as well as the Hamiltonian dynamics associated to unstable M3 branes in 11D supergravity and obtain the corresponding effective world-volume description around their respective tachyon vacua. 
\end{abstract}
\section{Overview and Motivation}
The Carroll symmetry, that was introduced earlier in \cite{carroll1}-\cite{Bacry:1968zf} has been started gaining a lot of attention in the recent years due to its several remarkable aspects. For example, it seems to - (I) have connections to that with the so called BMS symmetry algebra \cite{Duval:2014lpa}-\cite{Duval:2014uva}, (II) be exhibiting a duality with the nonrelativistic symmetry group \cite{Duval:2014uoa}, (III) have connections to that with the warped conformal field theory \cite{Hofman:2014loa} and so on.

Starting from particle dynamics \cite{Bergshoeff:2014jla}-\cite{Kowalski-Glikman:2014paa}, in various recent analysis the notion of Carroll symmetries had been successfully extended to system of classical strings, $ p $ branes \cite{Cardona:2016ytk} as well as several other exotic systems \cite{Hartong:2015xda}-\cite{Basu:2018dub}. However, in spite of all these attempts some answers are still lacking in the literature and some interesting directions are yet to be explored. For example, one might wonder what is the corresponding Carroll dynamics of a membrane \cite{Sezgin:2002rt}-\cite{Kluson:2008mz} in M theory and how does the equivalence between different world-volume descriptions hold in the Carroll limit of generic MP branes.

Studying Carroll limit for membranes should be regarded as an interesting direction in itself (in contrast to those of strings or $ p $ branes \cite{Cardona:2016ytk} studied previously) as it is naturally coupled to the background 3-form gauge fields ($ \mathcal{C}_{MNP}(X) $) \cite{Bozhilov:2005ew}. More specifically, going back to the previous analysis of \cite{Cardona:2016ytk}, we see that there the discussions on $ Dp $ branes as well as strings are made without the inclusion of background fields. However, on the other hand, it is also well known that the coupling between Carroll particles to that with background gauge fields unveils non trivial dynamics even when considering the single particle dynamics \cite{Bergshoeff:2014jla}. This therefore should clearly provide enough motivation for exploring the corresponding scenario in case of extended objects like $ p $ branes. In other words, unlike in the previous examples \cite{Cardona:2016ytk}, one should expect a different dynamics to pop up in the Carroll limit of extended objects like strings and $ p $ branes. The purpose of the present article is therefore to address these issues considering specific examples of M2 as well as tachyonic M3 branes in M theory. A naive expectation is that the corresponding dynamics of D2 branes in type IIA theory would follow quite naturally through dimensional reduction of transverse M2 branes in 11D. The other important issue is to show explicitly that different formulations of M2 brane world volume theory \cite{Bozhilov:2005ew} essentially lead to one unique world-volume description in the Carroll limit. Our analysis further reveals that these different formulations of the world-volume action essentially come up with different constraint structure at the Hamiltonian level. This is indeed a nontrivial exercise where following some field redefinition we finally show the equivalence between different world-volume theories in the Carroll limit of M2 branes. 

The present article therefore intends to explore all these issues taking specific examples of M2 as well as M3 branes in M theory. We start our analysis in Section 2 by considering M2 brane world-volume theory and thereby taking its subsequent Carroll limit(s) \cite{Cardona:2016ytk}. We outline the solutions corresponding to Hamilton's dynamical equations considering two different embeddings for Carroll M2 branes propagating over 11D target space geometry. In the next Section 3, we start with the traditional Nambu-Goto (NG) action \cite{Bozhilov:2005ew} for M2 branes and show the equivalence between different world-volume descriptions in the respective Carroll limits. We also construct the world-volume description for \textit{generic} (stable) MP branes in the Carroll limit. In Section 4, we move beyond the stable membrane configurations and consider Carroll limit corresponding unstable M3 branes in M theory.  Finally, we conclude our analysis in Section 5.
\section{Carroll M2 branes I}
\subsection{Hamiltonian formulation}
We start considering M2 branes moving over a curved manifold (equipped with a metric $ g_{MN}(X) $) in the presence of background 3-form fluxes ($ \mathcal{C}_{MNP}(X) $) \cite{Bozhilov:2005ew},
\begin{eqnarray}
\mathcal{S}_{M2}=\int d^3 \xi \mathcal{L}_{(M2)}\label{e1}
\end{eqnarray}
where, the Lagrangian could be formally expressed as\footnote{At this stage it is noteworthy to mention that there are two other equivalent (or parallel) descriptions for M2 brane world volume action namely the Polyakov and the Nambu-Goto form \cite{Bozhilov:2005ew}. However, for the first part of the analysis we prefer to stick with (\ref{e1}) as this form of action does not introduce additional non linearity into the dynamics of the system.},
\begin{eqnarray}
\mathcal{L}_{(M2)} = \frac{1}{4\lambda^{0}}\left[\mathcal{G}_{00}-2\lambda^{j}\mathcal{G}_{0j}+\lambda^{i}\lambda^{j}\mathcal{G}_{ij} -(2 \lambda^{0}T_2)^2 \det \mathcal{G}_{ij}\right] +T_2 \mathcal{C}_{012}.\label{e2}
\end{eqnarray}

Here we introduce the pullback of the background fields on the world-volume of the membrane,
\begin{eqnarray}
\mathcal{G}_{mn}=g_{MN}(X)\partial_{m}X^M \partial_{n}X^N;~\mathcal{C}_{012}=C_{MNP}(X)\partial_{0}X^M \partial_{1}X^N\partial_{2}X^P
\end{eqnarray} 
where, $ X^M (M,N=0,..,10) $ are the target space coordinates and $ \xi^{m}(m=(0,i)=0,1,2) $ are the world-volume directions. Finally, $ \lambda^{m} $s are the Lagrange multipliers and $ T_2 $ is the tension associated with the M2 branes.

In order to take a consistent Carroll limit \cite{Cardona:2016ytk} one needs to replace velocities in terms of momenta in the original action (\ref{e1}). We find the conjugate momentum,
\begin{eqnarray}
\Pi_M=\frac{\partial \mathcal{L}}{\partial (\partial_{0}X^M)}=\frac{g_{MN}}{4\lambda^{0}}\left( \partial_{0}X^N -2 \lambda^{j}\partial_{j}X^N\right) +T_2 \mathcal{C}_{MNP}\partial_{1}X^N \partial_{2}X^P\label{e4}
\end{eqnarray}
which could be inverted to find,
\begin{eqnarray}
\partial_{0}X^M= 4 \lambda^{0}\Pi^{M}+2\lambda^{j}\partial_{j}X^M -4\lambda^{0}T_2 g^{MP}\mathcal{C}_{PQN}\partial_{1}X^Q \partial_{2}X^N.\label{e5}
\end{eqnarray}

Substituting (\ref{e5}) into (\ref{e2}) we find\footnote{Notice that there is no explicit RR 3-form in the Lagrangian (\ref{e6}) when expressed in terms of conjugate momentum ($ \Pi_{M} $).},
\begin{eqnarray}
\mathcal{L} _{(M2)}=\Pi_{M}\partial_{0}X^M -(4\lambda^{0})^{-1}\mathcal{H}-\zeta ~\Pi^{M}\Pi^{N}g_{MN}=\Pi_{M}\partial_{0}X^M - \mathcal{N}\mathcal{H}_{P}\label{e6}
\end{eqnarray}
where we introduce the following constraint (corresponding to the Lagrange multiplier $ \zeta $),
\begin{eqnarray}
\Gamma = \Pi^{M}\Pi^{N}g_{MN} \approx 0
\end{eqnarray}
that eventually results in a consistent dynamics near the Carroll limit of the M2 brane. 

Finally we have the following \textit{primary} Hamiltonian constraint for the system,
\begin{eqnarray}
\mathcal{H}_{P}&=& \frac{\lambda^{0}T_{2}^{2}}{\mathcal{N}}\det \mathcal{G}_{ij}-\frac{\lambda^{i}\lambda^{j}}{4\lambda^{0}\mathcal{N}}\partial_{i}X^M\partial_{j}X^N g_{MN}+\frac{\zeta}{\mathcal{N}} \Pi^{M}\Pi^{N}g_{MN} \nonumber\\
& =& \mathcal{H}_{can}+\bar{\lambda}\Gamma \approx 0\label{e7}
\end{eqnarray}
where, $ \mathcal{H}_{can} $ is the standard canonical Hamiltonian and $\bar{\lambda} $ is the so called Lagrange multiplier in the presence of primary constraint(s) ($ \Gamma $) such that,
\begin{eqnarray}
\mathcal{H}_{P} \approx \mathcal{H}_{can}.
\end{eqnarray}

Notice that unlike the previous example for $ Dp $ branes \cite{Cardona:2016ytk}, here we have only one diffeomorphism constraint namely the Hamiltonian (\ref{e7}). The other constraint proportional to $ \Pi . \partial X $ is not quite apparent in this setup. Notice that this a particular feature of the starting action (\ref{e2}). However, as we see in the subsequent analysis that the constraint $ \Pi . \partial X $ indeed appears as a part of the Nambu-Goto (NG) formulation (\ref{E66}) of M2 branes (see (\ref{E73})). Finally, using a trivial field redefinition, we show that these two formulations are essentially equivalent to each other.
\subsection{Membrane Carroll limit}
We introduce the M2 brane Carroll limit as,
\begin{eqnarray}
X^{\mu}=\frac{\mathfrak{x}^{\mu}}{\omega}~;~\Pi_{\mu}=\omega \pi_{\mu}~,(\mu =0,1,2)\nonumber\\
X^{I}=\mathfrak{x}^{I}~;~\Pi_{I}= \pi_{I}~,(I,J,K =3,..,10)\label{e8}
\end{eqnarray}
where, we set $ \omega \rightarrow \infty $ towards the end of our calculations.

In order to take consistent Carroll limit, it is also necessary to consider appropriate scaling limit for Lagrange multipliers ($ \lambda^{m} $, $ \zeta $, $ \mathcal{N} $) as well as the membrane tension ($ T_2 $),
\begin{eqnarray}
\lambda^{0}=\frac{\tilde{\lambda}^{0}}{\omega^{2}}~;~\lambda^{i}=\frac{\tilde{\lambda}^{i}}{\omega}~;~T_2=\omega \tilde{T}_{2}~;~\mathcal{N}=\tilde{\mathcal{N}}~;~\zeta = \frac{\tilde{\zeta}}{\omega^{2}}.\label{e9}
\end{eqnarray}

Using (\ref{e8}) and (\ref{e9}) we finally obtain the Lagrangian in the Carroll limit as,
\begin{eqnarray}
\mathcal{L}^{(M2)}_{C}=\pi_{M}\partial_{0}\mathfrak{x}^M-\tilde{\mathcal{N}}\mathfrak{H}^{(M2)}_{P}\label{e10}
\end{eqnarray}
where we define,
\begin{eqnarray}
\mathfrak{H}^{(M2)}_{P}&=&\tilde{\lambda}^{0}\tilde{T}_{2}^{2}\det \gamma_{ij}-\frac{\tilde{\lambda}^{i}\tilde{\lambda}^{j}}{4\tilde{\lambda}^{0}}\partial_{i}\mathfrak{x}^I\partial_{j}\mathfrak{x}^J g_{IJ} +\tilde{\zeta}\pi^{\mu}\pi^{\nu}g_{\mu \nu}\nonumber\\
&=& \mathfrak{H}_{can} + \tilde{\zeta}\pi^{\mu}\pi^{\nu}g_{\mu \nu} \approx 0
\label{e12}
\end{eqnarray}
as being the primary Hamiltonian constraint associated to the Carroll limit of M2 branes. Notice that here $\gamma_{ij}=g_{IJ}\partial_{i}\mathfrak{x}^I \partial_{j}\mathfrak{x}^J (i,j=1,2) $. 

The resulting equations of motion could be formally expressed as\footnote{Here, $ | \gamma |=\det \gamma_{ij} $. },
\begin{eqnarray}
\label{e13}
\partial_{0}\mathfrak{x}^{\mu} =2 \tilde{\mathcal{N}}\tilde{\zeta}\pi_{\nu}g^{\mu \nu}~;~\partial_{0}\pi^{\mu}=0~;~\partial_{0}\mathfrak{x}^{I} =0,\\
\partial_{0}\pi_{I}+\tilde{\mathcal{N}}\tilde{\lambda}^{0}\tilde{T}_{2}^{2}(| \gamma |\gamma^{ij}\partial_{I}g_{JK}\partial_{i}\mathfrak{x}^J \partial_{j}\mathfrak{x}^K-2\partial_{i}(| \gamma | \gamma^{ij}\partial_{j}\mathfrak{x}^J g_{IJ}))+\frac{\tilde{\mathcal{N}}\tilde{\lambda}^{i}\tilde{\lambda}^{j}}{2\tilde{\lambda}^{0}}\partial_{i}(\partial_{j}\mathfrak{x}^J g_{IJ})=0
\label{e14}
\end{eqnarray}
which thereby implies that the transverse coordinates freeze in time hence like in the case for free strings/p branes \cite{Cardona:2016ytk} the M2 brane does not move in its Carroll limit. However, on the other hand, the momenta ($ \pi_{I} $) conjugate to transverse fluctuations ($ \mathfrak{x}^{I} $) turn out to be non trivial functions of world-volume coordinates. This is a special feature associated with the dynamics in the Carroll limit where there is in general no connection between spatial momenta and velocities of a dynamical Carroll object \cite{Cardona:2016ytk}. 

We now proceed towards solving the above set of equations (\ref{e13})-(\ref{e14}). In order to solve these equations one needs to consider specific M2 brane embeddings for different choice of subspaces within the full 11D target space geometry \cite{Bozhilov:2006bi}. In the following, we illustrate the procedure with two specific choices of M2 brane embeddings.
\subsubsection{Example I}
Consider Carroll M2 branes moving in $ R \times S^4 $ subspace of full $ AdS_4 \times S^7 $ geometry\footnote{We set AdS length scale $ L=1 $.} \cite{Bozhilov:2006bi},
\begin{eqnarray}
ds^2 = -dt^2 +4\left(d\psi^{2}+\cos^2 \psi d\varphi^{2}_{1}+\frac{\sin^{2}\psi}{2}(d\varphi^{2}_{2}+d\varphi^{2}_{3}) \right) 
\end{eqnarray}
where we set the angle $ \theta = \theta_{0}=\frac{\pi}{4} $ for which the background RR forms in $ AdS_4 $ vanishes. 

To start with we consider following embeddings for Carroll M2 branes,
\begin{eqnarray}
\label{e16}
\mathfrak{x}^{0}&=&t(\xi^{m})=\Lambda^{0}~_{0}~\xi^{0} ,\\
\mathfrak{x}^{1}&=&\psi (\xi^{2}),\\
\mathfrak{x}^{2}&=&\varphi_{1}(\xi^{m})=\Lambda^{2}~_{0}~\xi^{0},\\
\mathfrak{x}^{3}&=&\varphi_{2}(\xi^{m})=\Lambda^{3}~_{i}~\xi^{i},\\
\mathfrak{x}^{4}&=&\varphi_{3}(\xi^{m})=\Lambda^{4}~_{i}~\xi^{i}~(i=1,2)
\label{e20}
\end{eqnarray}
where we introduce constant coefficients $ \Lambda^{M}~_{m} (M=0,1,2,3,4~;~m=0,1,2) $ carrying mixed indices those are eventually constrained by the Carroll Hamiltonian (\ref{e12}).

Substituting (\ref{e16})-(\ref{e20}) into (\ref{e13}) we find,
\begin{eqnarray}
\pi^{0}=(2 \tilde{\mathcal{N}}\tilde{\zeta})^{-1}\Lambda^{0}~_{0}~;~\pi^{1}=0~;~\pi^{2}=(2 \tilde{\mathcal{N}}\tilde{\zeta})^{-1}\Lambda^{2}~_{0}.
\end{eqnarray}
Therefore, we find that the Carroll M2 brane extended along $ \mathfrak{x}^{1}=\psi $ direction and rotating around $ \mathfrak{x}^{2}$ direction moves in time with constant energy ($ \pi^{0} $) and momentum ($  \pi^{2}$). Notice that the M2 brane also wraps around two of the isometry directions ($ \varphi_{2} $ and $ \varphi_{3} $) of $ S^4 $. Finally, it is noteworthy to mention that with the above ansatz (\ref{e16})-(\ref{e20}) one trivially satisfies the rest of the two equations in (\ref{e13}).

Our next task would be to substitute (\ref{e16})-(\ref{e20}) into (\ref{e14}) in order to solve for the corresponding transverse momenta ($ \pi^{I} (I,J=3,4)$),
\begin{eqnarray}
\partial_{0}\pi^{I}-8\tilde{\mathcal{N}}\tilde{\lambda}^{0}\tilde{T}_{2}^{2}\sum_{J \neq I}\Lambda^{J}~_{1} \left(\Lambda^{J}~_{1}\Lambda^{I}~_{2}-\Lambda^{I}~_{1}\Lambda^{J}~_{2} \right)\sin 2\psi  \psi'(\xi^{2}) \nonumber\\
+ \frac{\tilde{\mathcal{N}}\tilde{\lambda}^{2}\tilde{\lambda}^{i}}{\tilde{\lambda}^{0}}\Lambda^{I}~_{i}\cot\psi \psi'(\xi^{2})=0,
\end{eqnarray}
which upon integration yields,
\begin{eqnarray}
\pi^{I} (\xi^{0},\xi^{i})=\int \mathcal{F}^{I}(\xi^{2})d\xi^{0} +\mathcal{C}
\end{eqnarray}
where $ \mathcal{C} $ is the constant of integration together with,
\begin{eqnarray}
\mathcal{F}^{I}(\xi^{2})=8\tilde{\mathcal{N}}\tilde{\lambda}^{0}\tilde{T}_{2}^{2}\sum_{J \neq I}\Lambda^{J}~_{1} \left(\Lambda^{J}~_{1}\Lambda^{I}~_{2}-\Lambda^{I}~_{1}\Lambda^{J}~_{2} \right)\sin 2\psi  \psi'(\xi^{2}) \nonumber\\
- \frac{\tilde{\mathcal{N}}\tilde{\lambda}^{2}\tilde{\lambda}^{j}}{\tilde{\lambda}^{0}}\Lambda^{I}~_{j}\cot\psi \psi'(\xi^{2}).
\end{eqnarray}

Finally we note down the Hamiltonian constraint (\ref{e12}) which for the present case yields,
\begin{eqnarray}
4\tilde{\lambda}^{0}\tilde{T}_{2}^{2}\left(\Lambda^{3}~_{1}\Lambda^{4}~_{2}-\Lambda^{4}~_{1}\Lambda^{3}~_{2} \right)^{2}\sin^{2}\psi-\frac{\tilde{\lambda}^{i}\tilde{\lambda}^{j}}{2\tilde{\lambda}^{0}}\Lambda^{I}~_{i}\Lambda^{I}~_{j} \nonumber\\
+\frac{\tilde{\zeta}^{-1}}{4 \tilde{\mathcal{N}}^{2}\sin^2 \psi}(4 \cos^2 \psi (\Lambda^{2}~_{0})^2 -(\Lambda^{0}~_{0})^2)\approx 0.
\label{e25}
\end{eqnarray}
\subsubsection{Example II}
As a second example, we consider Carroll M2 branes propagating over some specified subspace of the full $ AdS_7 \times S^4 $ geometry \cite{Bozhilov:2006bi},
\begin{eqnarray}
ds^2 = - \cosh^{2}\varrho dt^2 +d\varrho^{2}+\frac{1}{2}\sinh^{2}\varrho (d\psi^{2}+ d\chi^{2})+\frac{1}{4}(d\alpha^{2}+\cos^{2}\alpha d\theta^{2}).
\end{eqnarray}

We choose to work with the following embedding for Carroll M2 branes namely,
\begin{eqnarray}
\label{e28}
\mathfrak{x}^{0}&=&t(\xi^{m})=\Lambda^{0}~_{0}~\xi^{0} ,\\
\mathfrak{x}^{1}&=&\varrho (\xi^{2}),\\
\mathfrak{x}^{2}&=&\psi(\xi^{m})=\Lambda^{2}~_{m}~\xi^{m},\\
\mathfrak{x}^{3}&=&\chi(\xi^{m})=\Lambda^{3}~_{i}~\xi^{i},\\
\mathfrak{x}^{4}&=&\alpha(\xi^{2})\\
\mathfrak{x}^{5}&=&\theta (\xi^{m})=\Lambda^{5}~_{i}~\xi^{i}~(m=0,i=0,1,2)
\label{e33}
\end{eqnarray}
such that the M2 brane extends along $ \varrho $ and $ \alpha $, wraps around the isometry directions ($ \psi $, $ \chi $ and $ \theta $) and propagates with constant energy and momentum,
\begin{eqnarray}
\pi^{0}=(2 \tilde{\mathcal{N}}\tilde{\zeta})^{-1}\Lambda^{0}~_{0}~;~\pi^{1}=0~;~\pi^{2}=(2 \tilde{\mathcal{N}}\tilde{\zeta})^{-1}\Lambda^{2}~_{0}.
\end{eqnarray}

Next, we substitute (\ref{e28})-(\ref{e33}) into (\ref{e14}) which finally yields the following set of equations
\begin{eqnarray}
\partial_{0}\pi^3 -\tilde{\mathcal{N}}\tilde{\lambda}^{0}\tilde{T}_{2}^{2}\Lambda^{5}~_{1}(\Lambda^{3}~_{2}\Lambda^{5}~_{1}-\Lambda^{3}~_{1}\Lambda^{5}~_{2})\cos\alpha \times ~~~~\nonumber\\
 (-\sin\alpha \alpha' +\coth\varrho \cos\alpha \varrho') 
+ \frac{\tilde{\mathcal{N}}\tilde{\lambda}^{2}\tilde{\lambda}^{i}}{\tilde{\lambda}^{0}}\Lambda^{3}~_{i}\coth\varrho  \varrho' =0,
\end{eqnarray}
\begin{eqnarray}
\partial_{0}\pi^4 +4\tilde{\mathcal{N}}\tilde{\lambda}^{0}\tilde{T}_{2}^{2}|\gamma |\gamma^{ij}\Lambda^{5}~_{i}\Lambda^{5}~_{j}\sin 2\alpha
+ \frac{\tilde{\mathcal{N}}\tilde{\lambda}^{0}\tilde{T}_{2}^{2}}{2}(\Lambda^{5}~_{1})^{2}\sin 2\alpha \alpha'^2 +\frac{\tilde{\mathcal{N}}(\tilde{\lambda}^{2})^2}{2\tilde{\lambda}^{0}}\alpha'' \nonumber\\
-\tilde{\mathcal{N}}\tilde{\lambda}^{0}\tilde{T}_{2}^{2}(\Lambda^{3}~_{1})^{2} \sin 2\varrho \alpha' \varrho' -  \frac{\tilde{\mathcal{N}}\tilde{\lambda}^{0}\tilde{T}_{2}^{2}}{2}((\Lambda^{5}~_{1})^{2}\cos^2 \alpha +2( \Lambda^{3}~_{1})^{2} \sinh^{2}\varrho)\alpha'' =0,
\end{eqnarray}
\begin{eqnarray}
\partial_{0}\pi^5 -2\tilde{\mathcal{N}}\tilde{\lambda}^{0}\tilde{T}_{2}^{2}\Lambda^{3}~_{1}(\Lambda^{3}~_{2}\Lambda^{5}~_{1}-\Lambda^{3}~_{1}\Lambda^{5}~_{2})\sinh\varrho \times ~~~~\nonumber\\
(\alpha ' \tan \alpha \sinh \varrho -\varrho ' \cosh \varrho ) - \frac{\tilde{\mathcal{N}}\tilde{\lambda}^{2}\tilde{\lambda}^{i}}{\tilde{\lambda}^{0}}\Lambda^{5}~_{i}\tan\alpha \alpha' =0,
\end{eqnarray}
which could be further integrated to obtain the transverse momenta as,
\begin{eqnarray}
\pi^{I}=\pi^{I}(\xi^{0},\xi^{2})
\end{eqnarray}
provided we know the explicit form of the functions $ \varrho = \varrho (\xi^{2}) $ and $ \alpha = \alpha (\xi^{2}) $.

Finally, we note down the Hamiltonian constraint (\ref{e12}) for the system
\begin{eqnarray}
\tilde{\lambda}^{0}\tilde{T}_{2}^{2}\left( \alpha '^2 \left((\Lambda^{5}~_{1})^2 \cos ^2 \alpha +2 (\Lambda^{3}~_{1})^2 \sinh ^2 \varrho \right)+2 \cos ^2 \alpha \sinh ^2\varrho (\Lambda^{3}~_{2}\Lambda^{5}~_{1}-\Lambda^{3}~_{1} \Lambda^{5}~_{2})^2\right) \nonumber\\
-\frac{\tilde{\lambda}^{i}\tilde{\lambda}^{j}}{\tilde{\lambda}^{0}}(2\Lambda^{3}~_{i}\Lambda^{3}~_{j}\sinh^2\varrho +\Lambda^{5}~_{i}\Lambda^{5}~_{j}\cos^2\alpha)-\frac{(\tilde{\lambda}^2)^2}{\tilde{\lambda}^0}\alpha'^2 \nonumber\\
+\frac{2\tilde{\zeta}^{-1}}{\tilde{\mathcal{N}}^{2}}(\sinh^2\varrho (\Lambda^{2}~_{0})^2 -2(\Lambda^{0}~_{0})^2)\approx 0.
\label{e39}
\end{eqnarray}
\subsection{Stringy Carroll limit}
In the previous Section we discussed membrane Carroll limit associated to M2 branes where we rescale the first three target space coordinates. The purpose of this Section is to explore the so called stringy Carroll limit \cite{Cardona:2016ytk} associated to M2 branes where we scale the first two target space coordinates and their corresponding conjugate momenta,
\begin{eqnarray}
X^{\mu}=\frac{\mathfrak{x}^{\mu}}{\omega}~;~\Pi_{\mu}=\omega \pi_{\mu}~,(\mu =0,1)\nonumber\\
X^{I}=\mathfrak{x}^{I}~;~\Pi_{I}= \pi_{I}~,(I,J,K=2,..,10)\label{40}
\end{eqnarray}
such that $ \omega $ is set to infinity at the end.

Substituting (\ref{40}) into (\ref{e6}) we find,
\begin{eqnarray}
\mathcal{L}^{(S)}_{C}=\pi_{M}\partial_{0}\mathfrak{x}^M-\tilde{\mathcal{N}}\mathfrak{H}^{(S)}_{P}\label{e41}
\end{eqnarray}
where we note down the primary Hamiltonian (constraint) associated to stringy Carroll limit as,
\begin{eqnarray}
\mathfrak{H}^{(S)}_{P}=\tilde{\lambda}^{0}\tilde{T}_{2}^{2}\det \gamma_{ij}-\frac{\tilde{\lambda}^{i}\tilde{\lambda}^{j}}{4\tilde{\lambda}^{0}}\partial_{i}\mathfrak{x}^I\partial_{j}\mathfrak{x}^J g_{IJ} +\tilde{\zeta}\pi^{\mu}\pi^{\nu}g_{\mu \nu} \approx 0
\label{e42}
\end{eqnarray}
which is structurally identical to that with (\ref{e12}) except for the fact that now we have a different scaling (\ref{40}) that leads to a different 2 $ \times $ 2 matrix structure $ \gamma_{ij} $. Our next task would be to outline the dynamics of M2 branes in the string Carroll limit considering two specific examples as before. 
\subsubsection{Example I}
Consider the first example where Carroll M2 branes are moving in $ R \times S^4 $ subspace of full $ AdS_4 \times S^7 $ geometry \cite{Bozhilov:2006bi},
\begin{eqnarray}
ds^2 = -dt^2 +4\left(d\psi^{2}+\cos^2 \psi d\varphi^{2}_{1}+\frac{\sin^{2}\psi}{2}(d\varphi^{2}_{2}+d\varphi^{2}_{3}) \right).
\end{eqnarray} 

We choose to work with the following embeddings for Carroll M2 branes,
\begin{eqnarray}
\label{e44}
\mathfrak{x}^{0}&=&t(\xi^{m})=\Lambda^{0}~_{0}~\xi^{0} ,\\
\mathfrak{x}^{1}&=&\varphi_{1}(\xi^{m})=\Lambda^{1}~_{0}~\xi^{0},\\
\mathfrak{x}^{2}&=&\psi (\xi^{2}),\\
\mathfrak{x}^{3}&=&\varphi_{2}(\xi^{m})=\Lambda^{3}~_{i}~\xi^{i},\\
\mathfrak{x}^{4}&=&\varphi_{3}(\xi^{m})=\Lambda^{4}~_{i}~\xi^{i}~(i=1,2)
\label{e48}
\end{eqnarray}
which leads to the following conserved energy-momentum
\begin{eqnarray}
\pi^{0}=(2 \tilde{\mathcal{N}}\tilde{\zeta})^{-1}\Lambda^{0}~_{0}~;~\pi^{1}=(2 \tilde{\mathcal{N}}\tilde{\zeta})^{-1}\Lambda^{1}~_{0}
\end{eqnarray}
associated to M2 branes. Therefore, with the above embedding (\ref{e44})-(\ref{e48}) one essentially describes Carroll M2 branes extended along $ \mathfrak{x}^{2}=\psi $ direction and rotating along one of isometry directions ($ \varphi_{1} $) with constant angular momentum $ \Lambda^{1}~_{0} $. Finally, like in the previous example, the M2 brane also wraps around two of the isometry directions ($ \varphi_{2} $ and $ \varphi_{3} $) of $ S^4 $ along with respective winding numbers.

The rest of the Hamilton's equations of motion could be formally expressed as
\begin{eqnarray}
\partial_{0}\pi^{2}
-4\tilde{\mathcal{N}}\tilde{\lambda}^{0}\tilde{T}_{2}^{2}((\Lambda^{3}~_{1})^2 +(\Lambda^{4}~_{1})^2)\sin\psi (\psi '' \sin \psi +2 \psi '^2 \cos \psi ) \nonumber\\
+\frac{\tilde{\mathcal{N}}\tilde{\lambda}^{0}\tilde{T}_{2}^{2}}{2}|\gamma |\sin 2\psi \sum_{i=1}^{2}\sum_{I=3}^{4}\gamma^{ij}\Lambda^{I}~_{i}\Lambda^{I}~_{j}+\frac{\tilde{\mathcal{N}}(\tilde{\lambda}^{2})^2}{2\tilde{\lambda}^{0}}\psi'' =0,
\end{eqnarray}
\begin{eqnarray}
\partial_{0}\pi^{I} +8\tilde{\mathcal{N}}\tilde{\lambda}^{0}\tilde{T}_{2}^{2}\psi ' \sin 2 \psi \sum_{J\neq I}\Lambda^{J}~_{1}(\Lambda^{I}~_{1} \Lambda^{J}~_{2}-\Lambda^{I}~_{2} \Lambda^{J}~_{1})+\frac{\tilde{\mathcal{N}}\tilde{\lambda}^{2}\tilde{\lambda}^{i}}{2\tilde{\lambda}^{0}}\Lambda^{I}~_{i}\cot\psi \psi '=0
\end{eqnarray}
where $ I,J=3,4 $. Therefore the Hamiltonian dynamics could be solved provided we know the embedding, $ \psi = \psi (\xi^{2}) $.

Finally, we note down the Hamiltonian constraint (\ref{e42})
\begin{eqnarray}
\mathfrak{H}^{(S)}_{P}=\tilde{\lambda}^{0}\tilde{T}_{2}^{2}(8 \left((\Lambda^{3}~_{1})^2 +(\Lambda^{4}~_{1})^2\right) \psi '^2 \sin ^2\psi +4 \sin ^4\psi (\Lambda^{3}~_{2} \Lambda^{4}~_{1}-\Lambda^{3}~_{1} \Lambda^{4}~_{2})^2)\nonumber\\
-\frac{\tilde{\lambda}^{i}\tilde{\lambda}^{j}}{2\tilde{\lambda}^{0}}(\Lambda^{3}~_{i}\Lambda^{3}~_{j} +\Lambda^{4}~_{i}\Lambda^{4}~_{j})\sin^2 \psi -\frac{(\tilde{\lambda}^2)^2}{\tilde{\lambda}^0}\psi'^2 +\frac{\tilde{\zeta}^{-1}}{4 \tilde{N}^2}(-(\Lambda^{0}~_{0})^2 +4 (\Lambda^{1}~_{0})^2 \cos^2 \psi)\approx 0.
\end{eqnarray}
\subsubsection{Example II}
In the second example, we consider Carroll M2 brane embeddings in some subspace of $ AdS_7 \times S^4 $ geometry \cite{Bozhilov:2006bi},
\begin{eqnarray}
ds^2 = - \cosh^{2}\varrho dt^2 +d\varrho^{2}+\frac{1}{2}\sinh^{2}\varrho (d\psi^{2}+ d\chi^{2})+\frac{1}{4}(d\alpha^{2}+\cos^{2}\alpha d\theta^{2})
\end{eqnarray}
where we choose to work with the following ansatz,
\begin{eqnarray}
\label{e54}
\mathfrak{x}^{0}&=&t(\xi^{m})=\Lambda^{0}~_{0}~\xi^{0} ,\\
\mathfrak{x}^{1}&=&\psi(\xi^{m})=\Lambda^{1}~_{m}~\xi^{m},\\
\mathfrak{x}^{2}&=&\varrho (\xi^{2}),\\
\mathfrak{x}^{3}&=&\chi(\xi^{m})=\Lambda^{3}~_{i}~\xi^{i},\\
\mathfrak{x}^{4}&=&\alpha(\xi^{2})\\
\mathfrak{x}^{5}&=&\theta (\xi^{m})=\Lambda^{5}~_{i}~\xi^{i}~(m=0,i=0,1,2).
\label{e59}
\end{eqnarray}

Like in the previous example, a straightforward analysis reveals, 
\begin{eqnarray}
\pi^{0}=(2 \tilde{\mathcal{N}}\tilde{\zeta})^{-1}\Lambda^{0}~_{0}~;~\pi^{1}=(2 \tilde{\mathcal{N}}\tilde{\zeta})^{-1}\Lambda^{1}~_{0}.
\end{eqnarray}

The rest of the Hamilton's equations could be formally expressed as
\begin{eqnarray}
\partial_{0}\pi^{2}+\frac{\tilde{\mathcal{N}}\tilde{\lambda}^0 \tilde{T}^{2}_{2}}{2}| \gamma | \gamma^{ij}\sinh 2\varrho (\Lambda^{1}~_{i}\Lambda^{1}~_{j}+\Lambda^{3}~_{i}\Lambda^{3}~_{j})+\frac{\tilde{\mathcal{N}}(\tilde{\lambda}^{2})^2}{2\tilde{\lambda}^{0}}\varrho''-\frac{\tilde{\mathcal{N}}\tilde{\lambda}^0 \tilde{T}^{2}_{2}}{2}\times ~~~~~~~~~\nonumber\\
 \left(-(\Lambda^{5}~_{1})^2 \alpha ' \sin 2 \alpha \varrho '+\varrho '' \left((\Lambda^{5}~_{1})^2 \cos ^2\alpha +2 (\Lambda^{3}~_{1})^2 \sinh ^2\varrho \right)+2 (\Lambda^{3}~_{1})^2 \varrho '^2 \sinh 2 \varrho \right)=0,
\end{eqnarray}
\begin{eqnarray}
\partial_{0}\pi^{3}+\tilde{\mathcal{N}}\tilde{\lambda}^0 \tilde{T}^{2}_{2}\Lambda^{5}~_{1}\cos \alpha (\Lambda^{3}~_{2} \Lambda^{5}~_{1}-\Lambda^{3}~_{1} \Lambda^{5}~_{2}) \left(\alpha ' \sin \alpha -\cos \alpha \varrho ' \coth \varrho\right)\nonumber\\
+ \frac{\tilde{\mathcal{N}}\tilde{\lambda}^{2}\tilde{\lambda}^{i}}{\tilde{\lambda}^{0}}\Lambda^{3}~_{i}\coth\varrho  \varrho' =0,
\end{eqnarray}
\begin{eqnarray}
\partial_{0}\pi^{4}+4\tilde{\mathcal{N}}\tilde{\lambda}^{0}\tilde{T}_{2}^{2}|\gamma |\gamma^{ij}\Lambda^{5}~_{i}\Lambda^{5}~_{j}\sin 2\alpha+\frac{\tilde{\mathcal{N}}(\tilde{\lambda}^{2})^2}{2\tilde{\lambda}^{0}}\alpha'' -\frac{\tilde{\mathcal{N}}\tilde{\lambda}^0 \tilde{T}^{2}_{2}}{2}\times ~~~~~~~~~~~~~~\nonumber\\
 \left(\alpha '' \left((\Lambda^{5}~_{1})^2 \cos ^2\alpha +2 (\Lambda^{3}~_{1})^2 \sinh ^2\varrho \right)+2 (\Lambda^{3}~_{1})^2 \alpha ' \varrho ' \sinh 2 \varrho -(\Lambda^{5}~_{1})^2 \alpha '^2 \sin 2 \alpha \right)=0,
\end{eqnarray}
\begin{eqnarray}
\partial_{0}\pi^{5}+2\tilde{\mathcal{N}}\tilde{\lambda}^0 \tilde{T}^{2}_{2} \Lambda^{3}~_{1} \sinh \varrho (\Lambda^{3}~_{1} \Lambda^{5}~_{2}-\Lambda^{3}~_{2} \Lambda^{5}~_{1}) \left(\alpha ' \tan \alpha  \sinh \varrho -\varrho ' \cosh \varrho\right)\nonumber\\
-\frac{\tilde{\mathcal{N}}\tilde{\lambda}^{2}\tilde{\lambda}^{i}}{\tilde{\lambda}^{0}}\Lambda^{5}~_{i}\tan\alpha  \alpha' =0.
\end{eqnarray}

Finally, we note down the Hamiltonian constraint (\ref{e42}) for the configuration,
\begin{eqnarray}
\mathfrak{H}^{(S)}_{P}=\frac{\tilde{\lambda}^{0}\tilde{T}_{2}^{2}}{16} \left(\alpha '^2+4 \varrho '^2\right) \left((\Lambda^{5}~_{1})^2 \cos ^2\alpha +2 (\Lambda^{3}~_{1})^2 \sinh ^2\varrho \right)-\frac{(\tilde{\lambda}^2)^2}{4\tilde{\lambda}^0}\left( \varrho'^2 +\frac{\alpha'^2}{4}\right) \nonumber\\
+\frac{\tilde{\lambda}^{0}\tilde{T}_{2}^{2}}{8} \cos ^2\alpha \sinh ^2\varrho (\Lambda^{3}~_{2} \Lambda^{5}~_{1}-\Lambda^{3}~_{1} \Lambda^{5}~_{2})^2-\frac{\tilde{\lambda}^{i}\tilde{\lambda}^{j}}{8\tilde{\lambda}^{0}}(\sinh^2\varrho\Lambda^{3}~_{i}\Lambda^{3}~_{j} +\frac{1}{2}\cos^2 \alpha \Lambda^{5}~_{i}\Lambda^{5}~_{j}) \nonumber\\
+\frac{\tilde{\zeta}^{-1}}{4 \tilde{N}^2}(-(\Lambda^{0}~_{0})^2 +\frac{1}{2} (\Lambda^{1}~_{0})^2 \sinh^2 \varrho)\approx 0.
\end{eqnarray}
\section{Carroll M2 branes II}
This is a parallel computation for M2 branes (in its Carroll limit) where we choose to work with the traditional Nambu-Goto (NG) action for the world-volume theory and explore the corresponding Hamiltonian dynamics. We show that the present formalism is equivalent to that with the previous world-volume description of M2 branes under certain trivial transformations on the world-volume d.o.f. in the Carroll limit.

We start with the NG world-volume theory of the following form \cite{Bozhilov:2005ew},
\begin{eqnarray}
\mathcal{S}_{M2}=-\tau_{2}\int d^{3}\xi \sqrt{-\det \mathfrak{a}}+\frac{\tau_{2}}{3!}\int d^{3}\xi \varepsilon^{mnp}\mathcal{C}_{mnp}
\label{E66}
\end{eqnarray}
where, we introduce the world-volume metric as,
\begin{eqnarray}
 \mathfrak{a}_{mn}=g_{MN}\partial_{m}X^M \partial_{n}X^N~;~m,n,p=(0,i)=0,1,2.
\end{eqnarray}
together with the short hand notation for RR three form,
\begin{eqnarray}
\mathcal{C}_{mnp}=\mathcal{C}_{MNP}(X)\partial_{m}X^{M}\partial_{n}X^N\partial_{p}X^P.
\end{eqnarray}

Next, we note down the corresponding conjugate momenta as,
\begin{eqnarray}
\pi_{M}=-\tau_{2}\sqrt{-\det \mathfrak{a}}(\mathfrak{a}^{-1})^{0m}g_{MN}\partial_{m}X^N+\frac{\tau_{2}}{2!}\varepsilon^{ij}\mathcal{C}_{MNP}\partial_{i}X^N \partial_{j}X^P.
\label{E69}
\end{eqnarray}

Using (\ref{E69}), it is straightforward to express the world-volume action,
\begin{eqnarray}
\mathcal{S}_{M2}=\int \pi_{M}\partial_{0}X^M + ``Constarints".
\end{eqnarray}

In order to check the constraints we first define,
\begin{eqnarray}
\bar{\pi}_{M}=\pi_M -\frac{\tau_{2}}{2!}\varepsilon^{ij}\mathcal{C}_{MNP}\partial_{i}X^N \partial_{j}X^P
\end{eqnarray}
which yields the following primary constraint,
\begin{eqnarray}
\mathcal{G}_{P}=g^{MN}\bar{\pi}_{M}\bar{\pi}_{N}+\tau^{2}_{2}\det \mathfrak{a}^{(2)}_{ij}\approx 0.
\label{E72}
\end{eqnarray}

The other primary constraint we note down is the following,
\begin{eqnarray}
\mathcal{G}_{i}=\bar{\pi}_{M}\partial_{i}X^M \approx 0.
\label{E73}
\end{eqnarray}

Combining (\ref{E72}) and (\ref{E73}), we finally note down the M2 brane action as,
\begin{eqnarray}
\mathcal{S}_{M2}=\int \bar{\pi}_{M}\partial_{0}X^M +\frac{\tau_2}{3!}\varepsilon^{mnp}\mathcal{C}_{mnp}+\mathfrak{g}\mathcal{G}_{P}+\mathfrak{g}^{i}\mathcal{G}_{i}
\end{eqnarray}
where, $ \mathfrak{g} $ and $ \mathfrak{g}^{i} $ are the so called Lagrange multipliers.

Next we consider the so called \textit{stringy} Carroll limit associated to M2 branes,
\begin{eqnarray}
X^{\mu}=\frac{\mathfrak{x}^{\mu}}{\omega}~;~\bar{\pi}_{\mu}=\omega \tilde{\pi}_{\mu}~,(\mu =0,1)\nonumber\\
X^{I}=\mathfrak{x}^{I}~;~\bar{\pi}_{I}= \tilde{\pi}_{I}~,(I,J,K=2,..,10)
\end{eqnarray}
together with the following scaling,
\begin{eqnarray}
\mathfrak{g}=\frac{\tilde{\mathfrak{g}}}{\omega^{2}}~;~\mathfrak{g}^{i}=\tilde{\mathfrak{g}}^{i}~;~\tau_{2}=\omega \tilde{\tau}_{2}~;~\mathcal{C}_{MNP}=\omega^{-1}\tilde{\mathcal{C}}_{MNP}
\end{eqnarray}
which finally leads to the following Carroll M2 brane action,
\begin{eqnarray}
\tilde{\mathcal{S}}_{M2}=\int \tilde{\pi}_{M}\partial_{0}\mathfrak{x}^M+\frac{\tilde{\tau}_2}{3!}\varepsilon^{mnp}\tilde{\mathcal{C}}_{IJK}\partial_{m}\mathfrak{x}^I\partial_{n}\mathfrak{x}^J\partial_{p}\mathfrak{x}^K+\tilde{\mathfrak{g}}g^{\mu \nu}\tilde{\pi}_{\mu}\tilde{\pi}_{\nu}+\tilde{\mathfrak{g}}\tilde{\tau_2}^{2}\det\mathfrak{\gamma}_{ij}+\tilde{\mathfrak{g}}^{i}\tilde{\pi}_{M}\partial_{i}\mathfrak{x}^M.
\label{E77}
\end{eqnarray}

In order to check equivalence between (\ref{E77}) and (\ref{e41}) we consider the following field transformations,
\begin{eqnarray}
\tilde{\pi}_{\mu}=\tilde{\tilde{\pi}}_{\mu}~;~\tilde{\pi}_{I}=\tilde{\tilde{\pi}}_{I}-\frac{\tilde{\tau}_2}{2!}\varepsilon^{jk}\tilde{\mathcal{C}}_{IJK}\partial_{j}\mathfrak{x}^J\partial_{k}\mathfrak{x}^{K}\equiv \tilde{\tilde{\pi}}_{I}-\upsilon_{I}.\label{E78}
\end{eqnarray}

Substituting (\ref{E78}) into (\ref{E77}) we find,
\begin{eqnarray}
\tilde{\mathcal{S}}_{M2}=\int \tilde{\tilde{\pi}}_{M}\partial_{0}\mathfrak{x}^M+\tilde{\mathfrak{g}}g^{\mu \nu}\tilde{\tilde{\pi}}_{\mu}\tilde{\tilde{\pi}}_{\nu}+\tilde{\mathfrak{g}}\tilde{\tau_2}^{2}\det\mathfrak{\gamma}_{ij}+\tilde{\mathfrak{g}}^{i} \tilde{\tilde{\pi}}_{I}\partial_{i}\mathfrak{x}^I
\label{E79}
\end{eqnarray}
where we impose the following constraints namely,
\begin{eqnarray}
\tilde{\tilde{\pi}}_{\mu} \partial_{i}\mathfrak{x}^{\mu}\approx 0~;~\upsilon_{I} \partial_{i}\mathfrak{x}^I\approx 0.
\label{E80}
\end{eqnarray}

Therefore (\ref{E79}) is equivalent to (\ref{e41}) provided we set,
\begin{eqnarray}
\tilde{\tilde{\pi}}_{I}=\tilde{\mathfrak{g}}^{j}g_{IJ}\partial_j\mathfrak{x}^J
\end{eqnarray}
and identify the new set of Lagrange multipliers to that with the old set of Lagrange multipliers with proper signature and scaling.

The equations of motion that readily follow from (\ref{E77}) could be formally expressed as,
\begin{eqnarray}
\partial_0 \mathfrak{x}^{\mu}+2\tilde{\mathfrak{g}}g^{\mu \nu}\tilde{\pi}_{\nu}+\tilde{\mathfrak{g}}^{i}\partial_{i}\mathfrak{x}^{\mu}&=&0\\
\partial_0 \mathfrak{x}^{I}+\tilde{\mathfrak{g}}^{i}\partial_{i}\mathfrak{x}^{I}&=&0\\
\partial_{0}\tilde{\pi}_{\mu}+\tilde{\mathfrak{g}}^{i}\partial_{i}\tilde{\pi}_{\mu}&=&0\\
\partial_{0}\tilde{\pi}_{I}+\frac{\tilde{\tau_2}}{2!}\varepsilon^{mnp}\partial_{m}(\tilde{\mathcal{C}}_{IJK}\partial_n \mathfrak{x}^J \partial_p\mathfrak{x}^K)+2 \partial_{i}(\det \gamma_{ij}\gamma^{ij}g_{IJ}\partial_j \mathfrak{x}^J)+\tilde{\mathfrak{g}}^i \partial_i \tilde{\pi}_{I}&=&0
\end{eqnarray}
along with the following constraints,
\begin{eqnarray}
g^{\mu \nu}\tilde{\pi}_{\mu}\tilde{\pi}_{\nu}+\tilde{\tau}^2_2 \det \gamma_{ij}& \approx &0\nonumber\\
\tilde{\pi}_{M}\partial_{i}\mathfrak{x}^M & \approx & 0.
\end{eqnarray}
\textbf{Note added:} The above analysis could be generalized for generic MP ($ P>2 $ or 5) branes whose world-volume theory could be formally expressed as\footnote{Here, $ \mathcal{C}^{(P+1)} $ could be thought as being that of the \emph{composite} higher form (made out of the 3 form ($ \mathcal{C}^{(3)} $) and the corresponding field strength 4 form $ G^{(4)} (=d\mathcal{C}^{(3)}) $) in 11D SUGRA \cite{Cremmer:1978km}-\cite{Bagger:2012jb}. For example, in case of $P=5 $ one might think of a composite 6 form of the type, $\mathcal{C}^{(6)}= \star dG^{(4)}- \mathcal{C}^{(3)}\wedge \mathcal{C}^{(3)}$.},
\begin{eqnarray}
\mathcal{S}_{MP}=-\tau_{P}\int d^{P+1}\xi \sqrt{-\det \mathfrak{a}}+\tau_P \int\mathcal{C}^{(P+1)}
\label{E87}
\end{eqnarray}
where,
\begin{eqnarray}
\mathfrak{a}_{mn}&=&g_{MN}\partial_{m}X^M \partial_{n}X^N~;~m,n=0,..,P\\
\mathcal{C}^{(P+1)}&=&\frac{1}{(P+1)!}\mathcal{C}_{M_0..M_P}dX^{M_0}\wedge ..\wedge dX^{M_P}.
\end{eqnarray}

The corresponding conjugate momentum is given by,
\begin{eqnarray}
\pi_{M_0}=-\tau_{P}\sqrt{-\det \mathfrak{a}}(\mathfrak{a}^{-1})^{0m}g_{M_0N}\partial_{m}X^N+\frac{\tau_{P}}{P!}\varepsilon^{i_1..i_P}\mathcal{C}_{M_0M_1..M_P}\partial_{i_1}X^{M_1}.. \partial_{i_P}X^{M_P}.
\label{E90}
\end{eqnarray}

Using (\ref{E90}), it is now straightforward to express the world-volume action (\ref{E87}) as
\begin{eqnarray}
\mathcal{S}_{MP}=\int \hat{\pi}_{M_0} \partial_0 X^{M_0}+\frac{\tau_{P}}{(P+1)!}\varepsilon^{m_0..m_P}\mathcal{C}_{m_0..m_P}+\lambda \Phi_{P} +\lambda^{i}\Phi_{i}
\end{eqnarray}
where we introduce primary constraints,
\begin{eqnarray}
\Phi_{P} &=& g^{MN}\hat{\pi}_{M}\hat{\pi}_{N}+\tau_P^2 \det \mathfrak{a}_{ij}^{(P)}\approx 0\\
\Phi_{i} &=&\hat{\pi}_{M}\partial_{i}X^M \approx 0.
\end{eqnarray}
constructed out of modified canonical momentum,
\begin{eqnarray}
\hat{\pi}_{M_0}=\pi_{M_0}-\frac{\tau_{P}}{P!}\varepsilon^{i_1..i_P}\mathcal{C}_{M_0M_1..M_P}\partial_{i_1}X^{M_1}.. \partial_{i_P}X^{M_P}.
\end{eqnarray}

Next we introduce stringy Carroll limit as,
\begin{eqnarray}
X^{\mu}=\frac{\mathfrak{x}^{\mu}}{\omega}~;~\hat{\pi}_{\mu}=\omega \tilde{\pi}_{\mu}~,(\mu =0,1)\nonumber\\
X^{I}=\mathfrak{x}^{I}~;~\hat{\pi}_{I}= \tilde{\pi}_{I}~,(I=2,..,10)
\end{eqnarray}
which leads to the generic Carroll MP brane action,
\begin{eqnarray}
\tilde{\mathcal{S}}_{MP}=\int \tilde{\pi}_{M}\partial_{0}\mathfrak{x}^M+\frac{\tilde{\tau}_P}{(P+1)!}\varepsilon^{m_0..m_P}\tilde{\mathcal{C}}_{I_0..I_P}\partial_{m_0}\mathfrak{x}^{I_{0}}..\partial_{m_P}\mathfrak{x}^{I_P}+\tilde{\mathfrak{g}}g^{\mu \nu}\tilde{\pi}_{\mu}\tilde{\pi}_{\nu}\nonumber\\
+\tilde{\mathfrak{g}}\tilde{\tau_P}^{2}\det\mathfrak{\gamma}_{ij}+\tilde{\mathfrak{g}}^{i}\tilde{\pi}_{M}\partial_{i}\mathfrak{x}^M.
\label{E96}
\end{eqnarray}

Finally, with the following redefinition
\begin{eqnarray}
\tilde{\pi}_{\mu}=\hat{\hat{\pi}}_{\mu}~;~\tilde{\pi}_{I}=\hat{\hat{\pi}}_I -\frac{\tilde{\tau}_P}{P!}\varepsilon^{i_1..i_P}\mathcal{C}_{II_1..I_P}\partial_{i_1}X^{I_1}.. \partial_{i_P}X^{I_P}
\end{eqnarray}
and imposing the constraint conditions (\ref{E80}) we arrive at the generic MP brane world-volume action,
\begin{eqnarray}
\tilde{\mathcal{S}}_{MP}=\int \hat{\hat{\pi}}_{M}\partial_{0}\mathfrak{x}^M+\tilde{\mathfrak{g}}g^{\mu \nu}\hat{\hat{\pi}}_{\mu}\hat{\hat{\pi}}_{\nu}
+\tilde{\mathfrak{g}}\tilde{\tau_P}^{2}\det\mathfrak{\gamma}_{ij}+\tilde{\mathfrak{g}}^{i}\hat{\hat{\pi}}_I \partial_{i}\mathfrak{x}^I.
\label{E98}
\end{eqnarray}
\section{Tachyonic M3 branes}
\subsection{World-volume theory }
The purpose of this Section is to explore the Carroll dynamics associated to unstable M3 branes \cite{Intriligator:2000pk}-\cite{Kluson:2008mz} propagating over M theory supergravity background and in particular to find an interpretation of the corresponding Hamiltonian dynamics in terms of stable lower dimensional objects in M theory. We start with the first proposal for the unstable M3 brane world-volume action. The corresponding world-volume action could be formally expressed as,
\begin{eqnarray}
\mathcal{S}_{M3}=\tau_{3}\int d^{4}\xi \mathcal{L}_{(M3)}+\tau_{3}\int V(T)  dT \wedge C^{(3)}
\end{eqnarray}
where the world-volume Lagrangian is given by,
\begin{eqnarray}
\mathcal{L}_{(M3)}&=&-V(T)\sqrt{-\det \mathfrak{A}}\\
 \mathfrak{A}_{mn}&=&g_{MN}(X)\partial_{m}X^{M}\partial_{n}X^N+\partial_{m}T\partial_{n}T~;~m,n,p,q=(0,i)=0,1,2,3.
\end{eqnarray}
Here, $ V(T) $ is the tachyon potential such that $ V(0)(=\tau_{3}) $ equals the M3 brane tension and $ C^{(3)} $ is the background RR three form. 

We choose to work with the following background RR three form,
\begin{eqnarray}
C^{(3)}=\frac{1}{3!}\varepsilon_{MNP}dX^M \wedge dX^N \wedge dX^P
\end{eqnarray} 
which finally yields the following world-volume action,
\begin{eqnarray}
\mathcal{S}_{M3}=-\tau_{3}\int d^{4}\xi V(T)\sqrt{-\det \mathfrak{A}}+\frac{\tau_{3}}{3!}\int d\xi^{q}d^ 3 \xi V(T)  \varepsilon^{mnp}\mathcal{C}_{mnp}\partial_{q}T.
\label{e70}
\end{eqnarray}

A straightforward computation reveals the corresponding conjugate momenta as,
\begin{eqnarray}
\wp_{M}=-\tau_{3}V(T)\sqrt{-\det \mathfrak{A}}(\mathfrak{A}^{-1})^{0m}g_{MN}\partial_{m}X^N+\tau_{3}\frac{V(T)}{2!}\varepsilon^{ijk}\varepsilon_{MNP}\partial_{i}X^N\partial_{j}X^P\partial_{k}T
\label{e72}
\end{eqnarray}
and
\begin{eqnarray}
\wp_{T}=-\tau_{3}V(T)\sqrt{-\det \mathfrak{A}}(\mathfrak{A}^{-1})^{0m}\partial_{m}T+\tau_{3}\frac{V(T)}{3!}\varepsilon^{ijk}\mathcal{C}_{ijk}
\label{e73}
\end{eqnarray}
where the indices $ i,j,k(=1,2,3) $ run along the spatial directions of the M3 brane world-volume coordinates.

Using (\ref{e72}) and (\ref{e73}) it is straightforward to find the corresponding canonical Hamiltonian density,
\begin{eqnarray}
\mathfrak{H}^{(M3)}_{can}=\wp_{M}\partial_{0}X^M +\wp_{T}\partial_{0}T-\tau_{3}\mathcal{L}_{(M3)}-\tau_{3}\frac{V(T)}{3!}\varepsilon^{mnp}\mathcal{C}_{mnp}\partial_{q}T=0.
\label{e74}
\end{eqnarray}

Like in the previous example, we further note down the following primary constraints
\begin{eqnarray}
\label{e75}
\mathcal{H}_{\tau}&=&\bar{\wp}_{M}\bar{\wp}_{N}g^{MN}+\bar{\wp}_{T}^2+\tau^{2}_{3}V^2 (T)\det \mathfrak{A}^{(3)}_{ij}\approx 0,\\
\mathcal{H}_{i}&=&\bar{\wp}_{M}\partial_i X^M +\bar{\wp}_{T}\partial_{i}T \approx 0,
\label{e76}
\end{eqnarray}
where we introduce new canonical momenta as,
\begin{eqnarray}
\bar{\wp}_{M}&=&\wp_{M}-\tau_{3}\frac{V(T)}{2!}\varepsilon^{ijk}\varepsilon_{MNP}\partial_{i}X^N\partial_{j}X^P\partial_{k}T\\
\bar{\wp}_{T}&=&\wp_{T}-\tau_{3}\frac{V(T)}{3!}\varepsilon^{ijk}\mathcal{C}_{ijk}.
\end{eqnarray}

Using (\ref{e75}) and (\ref{e76}) we finally note down the unstable M3 brane action (\ref{e70}) as,
\begin{eqnarray}
\mathcal{S}_{M3}=\int d^{4}\xi   ~\Re (\bar{\wp}_{M}, X^M; \bar{\wp}_{T}, T)
\end{eqnarray}
where we introduce the function,
\begin{eqnarray}
\Re (\bar{\wp}_{M}, X^M; \bar{\wp}_{T}, T)=\bar{\wp}_{M}\partial_{0}X^M+\bar{\wp}_{T}\partial_{0}T+\tau_{3}\frac{V(T)}{2!}\varepsilon^{ijk}\varepsilon_{MNP}\partial_{i}X^N\partial_{j}X^P\partial_{k}T\partial_{0}X^M\nonumber\\
+\tau_{3}\frac{V(T)}{3!}\varepsilon^{ijk}\mathcal{C}_{ijk}\partial_{0}T +\ell \mathcal{H}_{\tau}+\ell^{i}\mathcal{H}_{i}
\label{e80}
\end{eqnarray}
together with the Lagrange multipliers $ \ell $ and $ \ell^{i} $.
\subsection{Carroll dynamics}
To start with, we introduce the \textit{stringy} Carroll limit corresponding to  M3 branes  as,
\begin{eqnarray}
X^{\mu}&=&\frac{\mathfrak{x}^{\mu}}{\omega}~;~ \bar{\wp}_{\mu}=\omega \bar{\pi}_{\mu}~,(\mu =0,1)\nonumber\\
X^{I}&=&\mathfrak{x}^{I}~;~\bar{\wp}_{I}=  \bar{\pi}_{I}~,(I,J , K=2,..,10)\nonumber\\
T&=&\frac{\mathfrak{t}}{\omega}~;~\bar{\wp}_{T}=  \omega \bar{\pi}_{\mathfrak{t}}~;~\tau_{3}=\omega \tilde{\tau}_{3}.
\label{e81}
\end{eqnarray}

Substituting (\ref{e81}) into (\ref{e80}) and taking the limit $ \omega \rightarrow \infty $ we find,
\begin{eqnarray}
\Re (\bar{\wp}_{M}, X^M; \bar{\wp}_{T}, T)=\bar{\pi}_{M}\partial_{0}\mathfrak{x}^M+\bar{\pi}_{\mathfrak{t}}\partial_{0}\mathfrak{t}+\frac{\tilde{\tau}_3^2}{2!}\varepsilon^{ijk}\varepsilon_{IJK}\partial_{i}\mathfrak{x}^J\partial_{j}\mathfrak{x}^K\partial_{k}\mathfrak{t}\partial_{0}\mathfrak{x}^I+\tilde{\ell}\bar{\pi}^2_{\mathfrak{t}}\nonumber\\
+\frac{\tilde{\tau}_3^2}{3!}\varepsilon^{ijk}\varepsilon_{IJK}\partial_{i}\mathfrak{x}^I\partial_{j}\mathfrak{x}^J\partial_{k}\mathfrak{x}^K\partial_{0}\mathfrak{t} +\tilde{\ell} \bar{\pi}_{\mu}\bar{\pi}_{\nu}g^{\mu \nu}+\tilde{\ell}\tilde{\tau}_3^4\det \mathfrak{a}^{(3)}_{ij}+\tilde{\ell}^{i}(\bar{\pi}_{M}\partial_{i}\mathfrak{x}^M+\bar{\pi}_{\mathfrak{t}}\partial_{i}\mathfrak{t})
\label{e82}
\end{eqnarray}
where we rescale the Lagrange multipliers as,
\begin{eqnarray}
\ell =\frac{\tilde{\ell}}{\omega^{2}}~;~\ell^{i} = \tilde{\ell}^{i}
\end{eqnarray}
together with the following identifications namely, $ \mathfrak{a}^{(3)}_{ij}=g_{IJ}\partial_{i}\mathfrak{x}^I \partial_{j}\mathfrak{x}^J $ and $ V(0)\sim \tilde{\tau}_{3} $.

The resulting equations of motion could be formally expressed as,
\begin{eqnarray}
\label{e84}
\partial_{0}\mathfrak{x}^{\mu}+2\tilde{\ell}\bar{\pi}_{\nu}g^{\mu \nu}+\tilde{\ell}^i \partial_i \mathfrak{x}^{\mu}&=&0\\
\partial_{0}\mathfrak{x}^{I}+\tilde{\ell}^i \partial_i \mathfrak{x}^{I}&=&0\\
\partial_{0}\mathfrak{t}+2 \tilde{\ell}\bar{\pi}_{\mathfrak{t}}+\tilde{\ell}^i \partial_i \mathfrak{t}&=&0\\
\partial_{0}\bar{\pi}_{\mu}+\tilde{\ell}^i \partial_i \bar{\pi}_{\mu}&=&0
\end{eqnarray}
\begin{eqnarray}
\partial_{0}\bar{\pi}_{I}+\tilde{\tau}_3^2\varepsilon^{ijk}\varepsilon_{IJK}\left[ \partial_{i}(\partial_{j}\mathfrak{x}^{[J}\partial_{0}\mathfrak{x}^{K]}\partial_{k}\mathfrak{t})+\frac{1}{2!}\partial_{0}(\partial_{i}\mathfrak{x}^{J}\partial_{j}\mathfrak{x}^{K}\partial_{k}\mathfrak{t})\right] +\tilde{\ell}^{i}\partial_{i}\bar{\pi}_{I}\nonumber\\
+\frac{\tilde{\tau}_3^2}{2!}\varepsilon^{ijk}\varepsilon_{IJK}\partial_{i}(\partial_{j}\mathfrak{x}^{J}\partial_{k}\mathfrak{x}^{K}\partial_{0}\mathfrak{t})+2\tilde{\ell}\tilde{\tau}_3^4\partial_{i}(\det \mathfrak{a}_{ij}\mathfrak{a}^{ij}g_{IJ}\partial_{j}\mathfrak{x}^J)=0
\end{eqnarray}
\begin{eqnarray}
\partial_{0}\bar{\pi}_{\mathfrak{t}}+\frac{\tilde{\tau}_3^2}{2!}\varepsilon^{ijk}\varepsilon_{IJK}\partial_{k}(\partial_{i}\mathfrak{x}^{J}\partial_{j}\mathfrak{x}^{K}\partial_{0}\mathfrak{x}^{I}) +\tilde{\ell}^{i}\partial_{i}\bar{\pi}_{\mathfrak{t}}
+\frac{\tilde{\tau}_3^2}{3!}\varepsilon^{ijk}\varepsilon_{IJK}\partial_{0}(\partial_{i}\mathfrak{x}^{I}\partial_{j}\mathfrak{x}^{J}\partial_{k}\mathfrak{x}^K)=0
\label{e89}
\end{eqnarray}
together with the following set of constraints namely,
\begin{eqnarray}
\bar{\pi}_{M}\partial_{i}\mathfrak{x}^M+\bar{\pi}_{\mathfrak{t}}\partial_{i}\mathfrak{t}&=&0\\
\bar{\pi}_{\mu}\bar{\pi}_{\nu}g^{\mu \nu}+\tilde{\tau}_3^4\det \mathfrak{a}_{ij}+\bar{\pi}^2_{\mathfrak{t}}&=&0.
\end{eqnarray}

Our next task would be to interpret the above dynamics (\ref{e84})-(\ref{e89}) at the tachyon vacuum $ \mathfrak{t}=\mathfrak{t}_{min} $. At the tachyon vacuum the world-volume action (\ref{e82}) reduces to, 
\begin{eqnarray}
\tilde{\mathcal{S}}_{M3}=\int \bar{\pi}_{M}\partial_{0}\mathfrak{x}^M +\tilde{\ell} \bar{\pi}_{\mu}\bar{\pi}_{\nu}g^{\mu \nu}+\tilde{\ell}\tilde{\tau}_3^4\det \mathfrak{a}^{(3)}_{ij}+\tilde{\ell}^{i}\bar{\pi}_{I}\partial_{i}\mathfrak{x}^I
\label{E124}
\end{eqnarray}
subjected to the vanishing of the tachyon momentum,
\begin{eqnarray}
\bar{\pi}_{\mathfrak{t}}=0.
\end{eqnarray}
and the constraints (\ref{E80}). The above action (\ref{E124}) essentially represents the effective world-volume theory that describes the dynamics of unstable M3 branes around their respective tachyon vacua.
\section{Summary and final remarks}
We now summarize our key results and conclude with some further remarks. The purpose of the present analysis was to explore the Carroll limit(s) in M theory. Taking specific examples of M2 branes as well as M3 branes we construct the corresponding world volume action in the Carroll limit and explore the associated Hamiltonian dynamics. The Hamilton's equations corresponding to transverse momenta turn out to be quite non trivial and is difficult to solve analytically. We also generalize our analysis by constructing world-volume theory corresponding to generic Mp branes moving over M theory supergravity backgrounds. We further show that different world-volume descriptions are equivalent in their respective Carroll limits. Finally, we turn our attention towards tachyonic M3 branes in M theory and their respective Carroll limits. We explore the corresponding membrane dynamics and obtain the world-volume theory at the tachyon vacua of the unstable Carroll membranes propagating over curved 11D M theory supergravity backgrounds.\\ \\ 
{\bf {Acknowledgements :}}
 The author is indebted to the authorities of IIT Roorkee for their unconditional support towards researches in basic sciences. \\\\ 



\begin{thebibliography}{99}
\bibitem{carroll1}J.M. L´evy-Leblond, Une nouvelle limite non-relativiste du group de Poincar´e, Ann. Inst.
Henri Poincar´e 3 (1965) 1.

\bibitem{carroll2} V.D. Sen Gupta, On an Analogue of the Galileo Group, Nuovo Cim. 54 (1966) 512.

\bibitem{Bacry:1968zf} 
  H.~Bacry and J.~Levy-Leblond,
  ``Possible kinematics,''
  J.\ Math.\ Phys.\  {\bf 9}, 1605 (1968).
  doi:10.1063/1.1664490
  
  \bibitem{Duval:2014lpa} 
  C.~Duval, G.~W.~Gibbons and P.~A.~Horvathy,
  ``Conformal Carroll groups,''
  J.\ Phys.\ A {\bf 47}, no. 33, 335204 (2014)
  doi:10.1088/1751-8113/47/33/335204
  [arXiv:1403.4213 [hep-th]].
  
  \bibitem{Duval:2014uva} 
  C.~Duval, G.~W.~Gibbons and P.~A.~Horvathy,
  ``Conformal Carroll groups and BMS symmetry,''
  Class.\ Quant.\ Grav.\  {\bf 31}, 092001 (2014)
  doi:10.1088/0264-9381/31/9/092001
  [arXiv:1402.5894 [gr-qc]].
  
  \bibitem{Duval:2014uoa} 
  C.~Duval, G.~W.~Gibbons, P.~A.~Horvathy and P.~M.~Zhang,
  ``Carroll versus Newton and Galilei: two dual non-Einsteinian concepts of time,''
  Class.\ Quant.\ Grav.\  {\bf 31}, 085016 (2014)
  doi:10.1088/0264-9381/31/8/085016
  [arXiv:1402.0657 [gr-qc]].
  
   \bibitem{Hofman:2014loa} 
  D.~M.~Hofman and B.~Rollier,
  ``Warped Conformal Field Theory as Lower Spin Gravity,''
  Nucl.\ Phys.\ B {\bf 897}, 1 (2015)
  doi:10.1016/j.nuclphysb.2015.05.011
  [arXiv:1411.0672 [hep-th]].

\bibitem{Bergshoeff:2014jla} 
  E.~Bergshoeff, J.~Gomis and G.~Longhi,
  ``Dynamics of Carroll Particles,''
  Class.\ Quant.\ Grav.\  {\bf 31}, no. 20, 205009 (2014)
  doi:10.1088/0264-9381/31/20/205009
  [arXiv:1405.2264 [hep-th]].
  
  \bibitem{Bergshoeff:2015wma} 
  E.~Bergshoeff, J.~Gomis and L.~Parra,
  ``The Symmetries of the Carroll Superparticle,''
  J.\ Phys.\ A {\bf 49}, no. 18, 185402 (2016)
  doi:10.1088/1751-8113/49/18/185402
  [arXiv:1503.06083 [hep-th]].
  
  \bibitem{Kowalski-Glikman:2014paa} 
  J.~Kowalski-Glikman and T.~Trześniewski,
  ``Deformed Carroll particle from 2+1 gravity,''
  Phys.\ Lett.\ B {\bf 737}, 267 (2014)
  doi:10.1016/j.physletb.2014.08.066
  [arXiv:1408.0154 [hep-th]].

  \bibitem{Cardona:2016ytk} 
  B.~Cardona, J.~Gomis and J.~M.~Pons,
  ``Dynamics of Carroll Strings,''
  JHEP {\bf 1607}, 050 (2016)
  doi:10.1007/JHEP07(2016)050
  [arXiv:1605.05483 [hep-th]].
  
  \bibitem{Hartong:2015xda} 
  J.~Hartong,
  ``Gauging the Carroll Algebra and Ultra-Relativistic Gravity,''
  JHEP {\bf 1508}, 069 (2015)
  doi:10.1007/JHEP08(2015)069
  [arXiv:1505.05011 [hep-th]].
  
  \bibitem{Barducci:2018wuj} 
  A.~Barducci, R.~Casalbuoni and J.~Gomis,
  ``Confined dynamical systems with Carroll and Galilei symmetries,''
  Phys.\ Rev.\ D {\bf 98}, no. 8, 085018 (2018)
  doi:10.1103/PhysRevD.98.085018
  [arXiv:1804.10495 [hep-th]].
  
  \bibitem{Barducci:2018thr} 
  A.~Barducci, R.~Casalbuoni and J.~Gomis,
  ``Vector SUSY models with Carroll or Galilei invariance,''
  Phys.\ Rev.\ D {\bf 99}, no. 4, 045016 (2019)
  doi:10.1103/PhysRevD.99.045016
  [arXiv:1811.12672 [hep-th]].
  
  \bibitem{Clark:2016qbj} 
  T.~E.~Clark and T.~ter Veldhuis,
  ``AdS-Carroll Branes,''
  J.\ Math.\ Phys.\  {\bf 57}, no. 11, 112303 (2016)
  doi:10.1063/1.4967969
  [arXiv:1605.05484 [hep-th]].
  
  \bibitem{Bergshoeff:2017btm} 
  E.~Bergshoeff, J.~Gomis, B.~Rollier, J.~Rosseel and T.~ter Veldhuis,
  ``Carroll versus Galilei Gravity,''
  JHEP {\bf 1703}, 165 (2017)
  doi:10.1007/JHEP03(2017)165
  [arXiv:1701.06156 [hep-th]].
  
  \bibitem{Duval:2017els} 
  C.~Duval, G.~W.~Gibbons, P.~A.~Horvathy and P.-M.~Zhang,
  ``Carroll symmetry of plane gravitational waves,''
  Class.\ Quant.\ Grav.\  {\bf 34}, no. 17, 175003 (2017)
  doi:10.1088/1361-6382/aa7f62
  [arXiv:1702.08284 [gr-qc]].
  
  \bibitem{Ciambelli:2018xat} 
  L.~Ciambelli, C.~Marteau, A.~C.~Petkou, P.~M.~Petropoulos and K.~Siampos,
  ``Covariant Galilean versus Carrollian hydrodynamics from relativistic fluids,''
  Class.\ Quant.\ Grav.\  {\bf 35}, no. 16, 165001 (2018)
  doi:10.1088/1361-6382/aacf1a
  [arXiv:1802.05286 [hep-th]].
  
  \bibitem{Basu:2018dub} 
  R.~Basu and U.~N.~Chowdhury,
  ``Dynamical structure of Carrollian Electrodynamics,''
  JHEP {\bf 1804}, 111 (2018)
  doi:10.1007/JHEP04(2018)111
  [arXiv:1802.09366 [hep-th]].
  
  \bibitem{Sezgin:2002rt} 
  E.~Sezgin and P.~Sundell,
  ``Massless higher spins and holography,''
  Nucl.\ Phys.\ B {\bf 644}, 303 (2002)
  Erratum: [Nucl.\ Phys.\ B {\bf 660}, 403 (2003)]
  doi:10.1016/S0550-3213(02)00739-3, 10.1016/S0550-3213(03)00267-0
  [hep-th/0205131].
  
  \bibitem{Alishahiha:2002sy} 
  M.~Alishahiha and M.~Ghasemkhani,
  ``Orbiting membranes in M theory on AdS(7) x S**4 background,''
  JHEP {\bf 0208}, 046 (2002)
  doi:10.1088/1126-6708/2002/08/046
  [hep-th/0206237].
  
  \bibitem{Alishahiha:2002fi} 
  M.~Alishahiha and A.~E.~Mosaffa,
  ``Circular semiclassical string solutions on confining AdS / CFT backgrounds,''
  JHEP {\bf 0210}, 060 (2002)
  doi:10.1088/1126-6708/2002/10/060
  [hep-th/0210122].
  
  \bibitem{Brugues:2004pj} 
  J.~Brugues, J.~Rojo and J.~G.~Russo,
  ``Non-perturbative states in type II superstring theory from classical spinning membranes,''
  Nucl.\ Phys.\ B {\bf 710}, 117 (2005)
  doi:10.1016/j.nuclphysb.2005.01.019
  [hep-th/0408174].
  
  \bibitem{Hartnoll:2002th} 
  S.~A.~Hartnoll and C.~Nunez,
  ``Rotating membranes on G(2) manifolds, logarithmic anomalous dimensions and N=1 duality,''
  JHEP {\bf 0302}, 049 (2003)
  doi:10.1088/1126-6708/2003/02/049
  [hep-th/0210218].
   
  \bibitem{Bozhilov:2005ew} 
  P.~Bozhilov,
  ``Membrane solutions in M-theory,''
  JHEP {\bf 0508}, 087 (2005)
  doi:10.1088/1126-6708/2005/08/087
  [hep-th/0507149].
  
  \bibitem{Bozhilov:2005xs} 
  P.~Bozhilov,
  ``Exact rotating membrane solutions on a G(2) manifold and their semiclassical limits,''
  JHEP {\bf 0603}, 001 (2006)
  doi:10.1088/1126-6708/2006/03/001
  [hep-th/0511253].
  
  \bibitem{Bozhilov:2003wr} 
  P.~Bozhilov,
  ``M2-brane solutions in AdS(7) x S**4,''
  JHEP {\bf 0310}, 032 (2003)
  doi:10.1088/1126-6708/2003/10/032
  [hep-th/0309215].
  
  \bibitem{Bozhilov:2006bi} 
  P.~Bozhilov and R.~C.~Rashkov,
  ``Magnon-like dispersion relation from M-theory,''
  Nucl.\ Phys.\ B {\bf 768}, 193 (2007)
  doi:10.1016/j.nuclphysb.2007.01.004
  [hep-th/0607116].
  
  \bibitem{Kim:2010ck} 
  J.~Kim, N.~Kim and J.~Hun Lee,
  ``Rotating Membranes in $AdS_4xM^{1,1,1}$,''
  JHEP {\bf 1003}, 122 (2010)
  doi:10.1007/JHEP03(2010)122
  [arXiv:1001.2902 [hep-th]].
  
  \bibitem{Axenides:2013lja} 
  M.~Axenides, E.~Floratos and G.~Linardopoulos,
  ``Stringy Membranes in AdS/CFT,''
  JHEP {\bf 1308}, 089 (2013)
  doi:10.1007/JHEP08(2013)089
  [arXiv:1306.0220 [hep-th]].
  
  \bibitem{Intriligator:2000pk} 
  K.~A.~Intriligator, M.~Kleban and J.~Kumar,
  ``Comments on unstable branes,''
  JHEP {\bf 0102}, 023 (2001)
  doi:10.1088/1126-6708/2001/02/023
  [hep-th/0101010].
  
  \bibitem{Houart:2000vm} 
  L.~Houart and Y.~Lozano,
  ``Brane descent relations in M theory,''
  Phys.\ Lett.\ B {\bf 479}, 299 (2000)
  doi:10.1016/S0370-2693(00)00317-8
  [hep-th/0001170].
  
  \bibitem{Kluson:2008mz} 
  J.~Kluson,
  ``Note About Unstable M3-brane Action,''
  Phys.\ Rev.\ D {\bf 79}, 026001 (2009)
  doi:10.1103/PhysRevD.79.026001
  [arXiv:0810.0585 [hep-th]].
  
  \bibitem{Cremmer:1978km} 
  E.~Cremmer, B.~Julia and J.~Scherk,
  ``Supergravity Theory in Eleven-Dimensions,''
  Phys.\ Lett.\  {\bf 76B}, 409 (1978).
  doi:10.1016/0370-2693(78)90894-8
  
  \bibitem{Bagger:2012jb} 
  J.~Bagger, N.~Lambert, S.~Mukhi and C.~Papageorgakis,
  ``Multiple Membranes in M-theory,''
  Phys.\ Rept.\  {\bf 527}, 1 (2013)
  doi:10.1016/j.physrep.2013.01.006
  [arXiv:1203.3546 [hep-th]].
 
\end{thebibliography}
\end{document}